\documentclass[conference]{IEEEtran}

\usepackage{graphicx} 
\usepackage{amsfonts} 
\usepackage{amsmath}
\usepackage{amssymb}
\usepackage{amsthm}
\usepackage{mathtools}
\usepackage{braket}
\usepackage{tikz}
\usepackage{algpseudocode}
\usepackage{algorithm}
\usepackage{float}
\usepackage{caption}
\usepackage{hyperref}
\usepackage[T1]{fontenc}

\DeclareCaptionFormat{algor}{%
  \hrulefill\par\offinterlineskip\vskip1pt%
    \textbf{#1#2}#3\offinterlineskip\hrulefill}
\DeclareCaptionStyle{algori}{singlelinecheck=off,format=algor,labelsep=space}
\usepackage{etoolbox}\AtBeginEnvironment{algorithmic}{\small}

\newtheorem{theorem}{Theorem}

\newtheorem{example}{Example}

\newenvironment{example*}
  {\addtocounter{example}{-1}\example}
  {\endexample}

\newcommand{\llbr}{[\![}
\newcommand{\rrbr}{]\!]}

\begin{document}

\title{GNarsil: Splitting Stabilizers into Gauges}
\makeatletter
\newcommand{\linebreakand}{%
  \end{@IEEEauthorhalign}
  \hfill\mbox{}\par
  \mbox{}\hfill\begin{@IEEEauthorhalign}
}
\makeatother

\author{
  \IEEEauthorblockN{Oskar Novak}
  \IEEEauthorblockA{\textit{Wyant College of Optical Sciences} \\
    \textit{University of Arizona}\\
    Tucson, USA 85721 \\
    E-mail: onovak@arizona.edu}
  \and
  \IEEEauthorblockN{Narayanan Rengaswamy}
  \IEEEauthorblockA{\textit{Department of Electrical and Computer Engineering} \\
    \textit{University of Arizona}\\
    Tucson, USA 85721 \\
    E-mail: narayananr@arizona.edu}
}


\maketitle

\begin{abstract}

Quantum subsystem codes have been shown to improve error-correction performance, ease the implementation of logical operations on codes, and make stabilizer measurements easier by decomposing stabilizers into smaller-weight gauge operators. 
In this paper, we present two algorithms that produce new subsystem codes from a ``seed'' CSS code.
They replace some stabilizers of a given CSS code with smaller-weight gauge operators that split the remaining stabilizers, while being compatible with the logical Pauli operators of the code. 
The algorithms recover the well-known Bacon-Shor code computationally as well as produce a new $\llbr 9,1,2,2 \rrbr$ rotated surface subsystem code with weight-$3$ gauges and weight-$4$ stabilizers.
We illustrate using a $\llbr 100,25,3 \rrbr$ subsystem hypergraph product (SHP) code that the algorithms can produce more efficient gauge operators than the closed-form expressions of the SHP construction.
However, we observe that the stabilizers of the lifted product quantum LDPC codes are more challenging to split into small-weight gauge operators.
Hence, we introduce the subsystem lifted product (SLP) code construction and develop a new $\llbr 775, 124, 20 \rrbr$ code from Tanner's classical quasi-cyclic LDPC code.
The code has high-weight stabilizers but all gauge operators that split stabilizers have weight $5$, except one.
In contrast, the LP stabilizer code from Tanner's code has parameters $\llbr 1054, 124, 20 \rrbr$.
This serves as a novel example of new subsystem codes that outperform stabilizer versions of them. 
Finally, based on our experiments, we share some general insights about non-locality's effects on the performance of splitting stabilizers into small-weight gauges.
\end{abstract}

\section{Introduction}

Quantum error correction is vital for quantum computers to achieve their full potential. 
The technique requires identifying the location of errors on a quantum computer without disturbing the delicate superposition states of the qubits involved in the computation. 
This is done by measuring stabilizers, which are quantum parity checks on different subsets of qubits, that help elucidate the locations of errors through an error syndrome. 
However, if measuring the stabilizers involves a high number of qubits, then the entangling measurements of the process pose the risk of creating additional errors.

Subsystem codes may help alleviate this issue~\cite{Kribs-prl05,Bacon-pra06}. 
These are codes with gauge operators or operators on virtual qubits that do not carry quantum information. 
If designed well, then the eigenvalue of a high-weight stabilizer can be obtained as a product of the eigenvalues of several lower-weight gauge operators that can be measured more easily.
Generally, these gauge operators form a non-abelian group, meaning that the order of measurement matters. 
The ordering can be chosen carefully during the process of \emph{gauge fixing}, where changes to the code can be made mid-computation to help ease measuring stabilizers or even implementing logical operations~\cite{nbreuckmann}. 
Recent work has even translated these gauge-fixing insights of subsystem codes into the ZX calculus~\cite{huang2023graphical}.  
There are several examples of subsystem codes in the literature. 
The prototypical example of a subsystem code is the Bacon-Shor code~\cite{Bacon-pra06}.
Most constructions have topological or geometrically local properties, which makes finding the gauge group an intuitive process~\cite{higgott2021subsystem}. 
Closed-form expressions exist for constructing generalized Bravyi-Bacon-Shor or Subsystem Hypergraph Product (SHP) Codes~\cite{li2020numerical}. 
There are also some computational methods for forming a set of gauge generators out of a set of Pauli operators via Gram-Schmidt orthogonalization~\cite{wilde2009logical}, and computational search methods for finding the optimal subsystem code from a set of two-qubit measurement operators~\cite{crosswhite2011automated}. 

However, there does not exist an algorithm that allows one to input a stabilizer code and derive a subsystem code directly from it, especially one that determines gauge operators that compose to form stabilizers of the input code. 
In recent years, there has been tremendous progress in constructing quantum low-density parity-check (QLDPC) stabilizer codes with optimal code parameters.
Such an algorithm can leverage these advances in stabilizer codes and potentially decompose their stabilizers into smaller-weight \emph{local} gauge checks.
For example, a good Lifted Product (LP) code can have stabilizer weights of $8$ or even larger than $10$, which involve long-range connections between qubits.
Hence, the LDPC property alone does not make these constructions practical.

In this work, we present two algorithms capable of deriving subsystem codes from a ``seed'' CSS stabilizer code and present non-trivial examples of subsystem codes found by our algorithms. 
The algorithms identify a $\llbr 9,1,2,2 \rrbr$ rotated surface subsystem code whose stabilizer weights are still $4$ but gauge weights are $3$, albeit with some non-locality.
In contrast, the subsystem surface code known in the literature~\cite{Bravyi-qic12} uses more qubits and has stabilizers of weight $6$.
Next, we demonstrate a modified SHP code that reduces the weight of gauge operators needed to produce a stabilizer compared to the closed-form expressions in~\cite{li2020numerical}.
We observe that it is in general difficult to decompose stabilizers of the LP construction into small-weight gauge operators.
Hence, we introduce an extension to the SHP construction that we call the \emph{Subsystem Lifted-Product (SLP)} codes, which can have superior parameters compared to the Lifted Product stabilizer code constructed from the same base matrix.
As an illustrative example, we produce a $\llbr 775, 124, 20 \rrbr$ SLP code from Tanner's classical quasi-cyclic LDPC code, whereas the corresponding LP code has parameters $\llbr 1054, 124, 20 \rrbr$.
The code has high-weight stabilizers but the gauge operators to produce the stabilizer are weight-$5$, except one of high weight.
It remains to be seen if the high-weight nature of stabilizers or some of the gauges is intrinsic to the SLP construction.

The paper is organized as follows:
Section~\ref{sec:background} introduces the necessary background to discuss the results in this work.
Section~\ref{sec:subsystem_binary_rep} reviews the binary symplectic representation of stabilizer and subsystem codes.
Section~\ref{sec:gnarsil_algorithms} provides our algorithms for splitting stabilizers into gauges.
Section~\ref{sec:examples} discusses some non-trivial examples found using our algorithms, including the SHP and SLP constructions.
Section~\ref{sec:conclusion} concludes the paper with thoughts about future work.


\section{Background}
\label{sec:background}

\subsection{Representing Pauli Operators as  Binary Vectors}
Given the vectors $ a=[a_{1},...,a_{n}]$, $ b=[b_{1},...,b_{n}]\in \mathbb{F}^{n}_{2}$, we define the Hermitian operator
\begin{equation}
    E(a,b) \coloneqq i^{a_{1}b_{1}}X^{a_{1}}Z^{b_{1}}\otimes...\otimes i^{a_{n}b_{n}}X^{a_{n}}Z^{b_{n}}, 
\end{equation}
where $X$ and $Z$ are the standard Pauli operators and $i \coloneqq \sqrt{-1}$.
We can define the $n$-qubit Pauli group $\mathcal{P}_{n}$ as
\begin{equation}
\begin{aligned}
    \mathcal{P}_{n} \coloneqq \left< i^{\alpha} E(a,b)\ \colon \ a,b \in \mathbb{F}^{n}_{2}; \alpha \in \{0,1,2,3\} \right>.
\end{aligned}
\end{equation}
This is also known as the Heisenberg-Weyl Group $HW_{N}, \ N=2^{n}$. 
The standard symplectic inner product in $\mathbb{F}^{2n}_{2}$ is defined as \cite{Rengaswamy-tqe20}:
\begin{align}
\left<[a,b],[a',b']\right>_{s} & \coloneqq a'b^{T}+b'a^{T} \ (\mathrm{mod} \ 2) \nonumber \\
\label{eq:symp_inn_pdt}
  &= [a,b]\ \mathrm{\boldsymbol{\Omega}}\ [a',b']^{T} \ (\mathrm{mod} \ 2),
\end{align}
where the matrix $\boldsymbol{\Omega}=\begin{bmatrix} 
  \boldsymbol{0} & \boldsymbol{I}_n \\
  \boldsymbol{I}_n & \boldsymbol{0}
  \end{bmatrix}$ is the symplectic form in $\mathbb{F}^{2n}_{2}$.
We notice that two operators  $E(a,b)$, $E(a',b')$ commute if and only if $\left<[a,b],[a',b']\right>_{s} =0$ (mod 2). Thus, we see that a vector $[a,b] \in \mathbb{F}^{2n}_{2} $ is isomorphic to an operator $E(a,b)$ via the map $\gamma: \mathcal{P}_{n}/\left<i^{\alpha} \boldsymbol{I}_{2^n}\right> \xrightarrow{} \mathbb{F}^{2n}_{2}$ defined by
\begin{equation}
\gamma(E(a,b)) \coloneqq [a,b].
\end{equation}
Thus, without loss of generality, we can represent $n$-qubit Pauli operators as binary vectors. 
The \textit{weight} of an $n$-qubit Pauli operator is the number of qubits on which it applies a non-identity Pauli operator, e.g., the operator $X_{1}\otimes X_{2} \otimes I_{3}$ can be described as $\begin{bsmallmatrix} 1 & 1 & 0, & 0 & 0 & 0 \end{bsmallmatrix}$, and its weight is $2$.

\subsection{Quantum Stabilizer Codes}

A stabilizer group $\mathcal{S}\in \mathcal{P}_{n}$ is an abelian subgroup of the Pauli group that does not contain $-I$. 
The corresponding \textit{stabilizer code} is defined by: 
\begin{equation} 
\mathcal{Q} \coloneqq \left\{ \ket{\psi} \in \mathbb{C}^{2^n} \colon S\ket{\psi}=\ket{\psi} \ \forall\ S \in \mathcal{S} \right\}.
\end{equation} 
A stabilizer code with $n$ physical qubits and $m$ independent stabilizer generators can encode $k=n-m$ logical qubits. 
The logical Pauli operators of the code come from the normalizer of $\mathcal{S}$ in $\mathcal{P}_{n}$, which is also its centralizer, defined as
\begin{equation}
\mathcal{N}(\mathcal{S}) \coloneqq \left\{ U\in \mathcal{P}_{n} \colon [U,S]=0 \ \forall\ S \in \mathcal{S} \right\},
\end{equation}
where $[U,S] \coloneqq US - SU$ is the commutator of $U$ and $S$.
Here, notice that $\mathcal{S} \subset \mathcal{N}(\mathcal{S})$. 
Finally, the \textit{minimum distance}, $d$, of the code is given by the minimum weight of any Pauli operator in $\mathcal{N}(\mathcal{S}) \setminus \mathcal{S}$, or the lowest weight of any logical operator. 
Thus, we denote the parameters of a quantum stabilizer code as $\llbr n,k,d \rrbr$. 
Finally, a CSS code is a stabilizer code whose stabilizer $\mathcal{S}=\left< \gamma^{-1}(\boldsymbol{H}_{X}),\gamma^{-1}(\boldsymbol{H}_{Z})\right>$, $\boldsymbol{H}_{X}\boldsymbol{H}_{Z}^{T}=\boldsymbol{0}$, where $\boldsymbol{H}_{X}$, $\boldsymbol{H}_{Z}$ are classical binary parity check matrices, and the map $\gamma^{-1}$ is applied to each row of $\boldsymbol{H}_{X}$, $\boldsymbol{H}_{Z}$.

\subsection{Subsystem Codes}

A subsystem code is a quantum error-correcting code that splits the code space $\mathcal{C}=A\otimes B$ into the logical subspace, $A$, and a gauge subspace, $B$, that does not carry any logical information~\cite{Kribs-prl05}. 
The Hilbert space of a subsystem code can be written as $\mathcal{H}=\mathcal{C} \oplus \mathcal{C}^{\perp}=A \otimes B \oplus \mathcal{C}^{\perp}$. 
The gauge subspace is supported on $r$ gauge qubits such that, given $n$ physical qubits and $m$ independent stabilizers, the number of logical qubits is $k=n-m-r$.  
Thus, a subsystem code with these parameters and code distance $d$ can be written as an $ \llbr n,k,r,d \rrbr$ subsystem code. 
A subsystem code is defined by its gauge group 
\begin{equation}
    \mathcal{G} \coloneqq \left< iI,\mathcal{S},X_{1}',Z_{1}', \ldots, X_{r}',Z_{r}' \right> \subset \mathcal{P}_{n},
\end{equation} 
where $X_{i}', Z_{i}'$ are the  Pauli  operators for the $i$-th gauge qubit, and $\mathcal{S}$ is the stabilizer group of the code. 

We can use $\mathcal{G}$ to define the stabilizers of the code: 
\begin{equation}
    \mathcal{S}\coloneqq\mathcal{Z}(\mathcal{G})=\mathcal{C}(\mathcal{G}) \cap \mathcal{G},
\end{equation} 
where $\mathcal{Z}(\mathcal{G})$ is the center of $\mathcal{G}$, and  $\mathcal{C}(\mathcal{G})$ is the centralizer of $\mathcal{G}$ in $\mathcal{P}_{n}$. 
We define the bare logical operators $\mathcal{L}_{b}$ as: 
\begin{equation}
\mathcal{L}_{b}\coloneqq\mathcal{C}(\mathcal{G}) \setminus \mathcal{G} = \mathcal{C}(\mathcal{G}) \setminus \mathcal{S}.
\end{equation} 
These are generated by $k$ pairs of anti-commuting Pauli operators such that $[\mathcal{G},\mathcal{L}_b]=0$. 
These operators only act non-trivially on the subspace $A$. We also have the dressed logical operators $\mathcal{L}$, $\mathcal{L}_{b} \subset \mathcal{L}$, where: 
\begin{equation} 
\mathcal{L}=\mathcal{C}(\mathcal{S}) \setminus \left< i\mathcal{I} \right>. 
\end{equation} 
In other words, the set of dressed logical operators is the set of bare logical operators multiplied by an operator from $\mathcal{G} \setminus \left< i\mathcal{I} \right>$.

\begin{figure}[h]
 \includegraphics[scale=0.4]{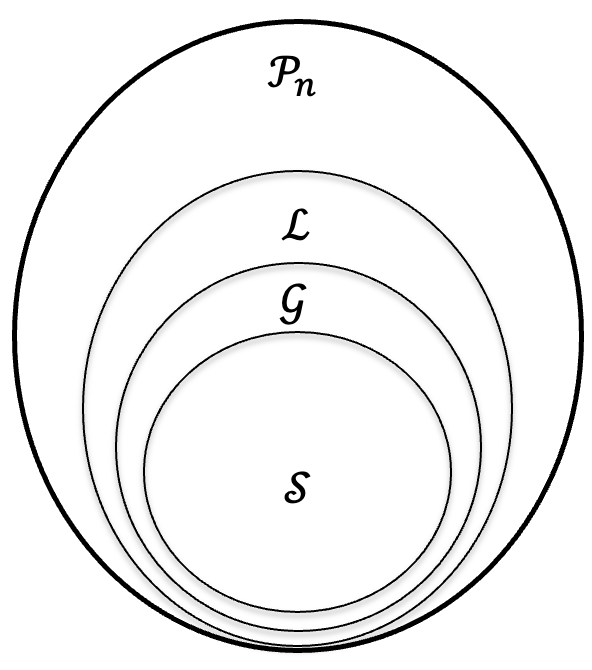}
\centering
\caption{A diagram describing the relation of the Gauge group $\mathcal{G}$ to the other important subsets of $\mathcal{P}_{n}$ for a subsystem code.}
\end{figure}

Finally, we can define the code distance $d$ such that 
\begin{equation}
    d \coloneqq \underset{P \in \mathcal{L}}{\min}\ |P|,
\end{equation} 
i.e., the minimum weight of a dressed logical operator.

\section{Binary Description of Subsystem Codes}
\label{sec:subsystem_binary_rep}

\subsection{Stabilizer Code Construction via Binary Matrices}

We can describe a stabilizer code as a $2n \times 2n$ matrix $\boldsymbol{U}$ which has the following form \cite{Aaronson-pra04,Dehaene-pra03,Rengaswamy-tqe20}:
\begin{equation} 
\label{eq:U_matrix}
\boldsymbol{U}=
\begin{bmatrix}
    \mathcal{L}_{X} \\
    \mathcal{S} \\
    \mathcal{L}_{Z} \\
    \mathcal{S}'
\end{bmatrix},
\end{equation}
where its rows $\left<\boldsymbol{u}_{1}, \ldots ,\boldsymbol{u}_{2n}\right>$ span $\mathbb{F}^{2n}_{2}$. 
We are overloading notation to represent both the group as well as a binary matrix whose rows represent the generators of the group.
Here, $\mathcal{L}_{X}$ and  $\mathcal{L}_{Z}$ are the binary matrices that represent the generators of the logical Pauli operators. 
The sub-matrix $\mathcal{S}'$, otherwise known as the ``destabilizer''~\cite{Aaronson-pra04}, is added so that $\boldsymbol{U}$ has full rank. For a stabilizer code, we choose $\mathcal{S}'$ such that:
\begin{equation}
\label{eq:U_condition}
\boldsymbol{U}\boldsymbol{\Omega}\boldsymbol{U}^T=\boldsymbol{\Omega} \ (\mathrm{mod} \ 2).
\end{equation}
This means that each row vector in $\mathcal{S}'$ is chosen so that it anti-commutes with only one stabilizer generator inside $\mathcal{S}$. 
In other words, $\boldsymbol{U}$ is a binary symplectic matrix~\cite{Dehaene-pra03}.

\begin{theorem} 
Let $\boldsymbol{U}$ be a binary matrix constructed as in~\eqref{eq:U_matrix}. 
If $\boldsymbol{U}$ satisfies~\eqref{eq:U_condition}, then $\boldsymbol{U}$ describes a valid stabilizer code.
\begin{IEEEproof}
\normalfont
The L.H.S. of~\eqref{eq:U_condition} is our definition of the symplectic inner product in~\eqref{eq:symp_inn_pdt} extended to a matrix. 
Since $\boldsymbol{\Omega}$ has a single non-zero element per row, each row of $\boldsymbol{U}$ anti-commutes with exactly one other row, which is $n$ rows below (or above) it. 
Such pairs of rows are called \textit{symplectic pairs}.
Since an $ \llbr n,k,d \rrbr$ stabilizer code 
has $n-k$ stabilizers, the symplectic pairs of rows in $\mathcal{L}_{X}$ must be in the rows of $\mathcal{L}_{Z}$. 
This means that the codes' logical operators commute with the operators represented as binary vectors in $\mathcal{S}$, $\mathcal{S}'$. 
Thus, satisfying~\eqref{eq:U_condition} verifies all necessary conditions for $\boldsymbol{U}$ to represent a valid stabilizer code.
\end{IEEEproof}
\end{theorem}

\subsection{Subsystem Code Construction via Binary Matrices}

To construct a subsystem code using the methods of the previous section, we can take a binary matrix $\boldsymbol{U}$ constructed as in \eqref{eq:U_matrix}, and choose $2r$ rows of it to become our gauge generators, and $\mathcal{L}_{X}$, $\mathcal{L}_{Z}$ become our bare logical operators. 

It should be noted that, generally, one can replace these $2r$ rows with other rows and still have a valid subsystem code, as long as $\boldsymbol{U}$ remains symplectic. 
Also, one is free to choose logical operators to become gauge operators. 

\begin{theorem} 
Let $\boldsymbol{U}$ be a $2n \times 2n$ binary matrix that has the construction of~\eqref{eq:U_matrix} except that  $2r$ of its rows have possibly been altered. We call this altered matrix $\boldsymbol{U}'$. 
Then $\boldsymbol{U}'$ represents a valid subsystem code if:
\begin{equation}
\label{eq:U'_matrix}\boldsymbol{U'}\boldsymbol{\Omega}\boldsymbol{U}'^T=\boldsymbol{\Omega} \ (\mathrm{mod} \ 2).
\end{equation}
\begin{IEEEproof}
\normalfont
Since only the $2r$ rows that have been altered are those promoted to gauge operators, the rest of the rows of the matrix on the R.H.S. of~\eqref{eq:U'_matrix} are identical to the rows in the same positions in~\eqref{eq:U_matrix}. 
This means that the bare logical operators only anti-commute with their respective symplectic pairs and commute with the operators that generate the gauge group. 
Furthermore, this also means that since the leftover stabilizers in $\mathcal{S}$ only anti-commute with the leftover operators in $\mathcal{S}'$, the operators of $\mathcal{S}$ indeed form $\mathcal{Z}(\mathcal{G})$. 
Finally, as long as the previous conditions are met, then \eqref{eq:U'_matrix} guarantees that $\boldsymbol{U}'$ represents a valid subsystem code.
\end{IEEEproof}
\end{theorem}
Appendix~\ref{sec:AppendixA} proves our algorithm's ability to find all suitable representations for the gauge generators.

\section{GNarsil Algorithms to Find Gauge Operators}
\label{sec:gnarsil_algorithms}

Here we will use \textit{gauge generators} to describe the non-stabilizer generators of the gauge group.
In practice, a subsystem code's utility comes from measuring the product of a set of lower-weight gauge operators that yields the same information as measuring a higher-weight stabilizer \cite{higgott2021subsystem}. 
Using the binary matrix constructions of the previous section, we can find low-weight gauge operators for CSS codes whose composition produces the code's stabilizers up to a remaining gauge operator that is not a gauge generator.
Let us call this remaining gauge operator the \emph{residual operator} for that stabilizer.
We refer to the minimum (Pauli) weight over all residual operators as the \textit{residual weight}.
For a gauge decomposition to be useful, the residual weight must be less than the weight of the decomposed stabilizer. 
To achieve this, we propose two algorithms. 
The first algorithm finds the set of $r$ gauge generators that decompose stabilizers with the least residual weight and then adds back stabilizers so that $n=k-r-m$ is still satisfied.  
We describe the first \emph{GNarsil} algorithm, which ``cuts'' stabilizers into gauges, in Algorithm \ref{alg:euclid}.

We also note that by bypassing the anti-commutation conditions, gauge operators that are not necessarily gauge generators may be found such that they decompose a stabilizer with lower residual weight. 
We also provide a second \emph{GNarsil} algorithm for this case in Algorithm \ref{alg:euclid2}.

\begin{algorithm}
\caption{\emph{GNarsil} 1: Gauge Generators to Split Stabilizers}
\label{alg:euclid}

\begin{algorithmic}[1]

\State \textbf{Input:} $\boldsymbol{U} = 
\begin{bmatrix}
    \mathcal{L}_{X} \\
    \mathcal{S} \\
    \mathcal{L}_{Z}\\
    \mathcal{S'}
\end{bmatrix}$, $\{ i_1, i_2, \ldots, i_r \} \subseteq \{ k+1, k+2, \ldots, n \}:$ indices of rows of $\boldsymbol{U}$ to be replaced with $X$-type gauge generators, desired Pauli weight $w$ for gauge generators

\State \textbf{Initialization:} 



$\mathcal{G}_X \gets \emptyset, \ \mathcal{G}_Z \gets \emptyset$

$\mathrm{validXGauges} \gets \emptyset, \ \mathrm{validZGauges} \gets \emptyset$



$\mathrm{maxSize} \gets $ max. number of gauge candidates to consider; 

$S_{\rm Xtargets} \gets$ indices of rows of $\boldsymbol{U}$ that are $X$-stabilizers; 

$S_{\rm Ztargets} \gets$ indices of rows of $\boldsymbol{U}$ that are $Z$-stabilizers; 

$\mathrm{gaugesPerStab} \gets $ \# of gauge generators per stabilizer; 

Xops $\gets$ weight-$w$ (row) vectors in $\{0,1\}^{n}$ appended with 

\hspace{11mm} $\boldsymbol{0}$ at the end to make length $2n$ (for $X$-gauges); 

Zops $\gets$ weight-$w$ (row) vectors in $\{0,1\}^{n}$ appended with 

\hspace{11mm} $\boldsymbol{0}$ at the front to make length $2n$ (for $Z$-gauges); 

$\boldsymbol{V}\gets \boldsymbol{U}(1:(n+k))$ (i.e., remove $\mathcal{S}'$ from $\boldsymbol{U}$) 

\Comment{Find $X$, $Z$ gauge generators}

\For{$i$ in 1:size(Xops)}
  
  \If{size(validXGauges) $\geq$ maxSize}
  \State \textbf{break}
  \EndIf
  
  \If{$\mathrm{Xops}(i)\, \boldsymbol{\Omega}\, \boldsymbol{V}^T = \boldsymbol{0}\ (\bmod\ 2)$ and \\%
      \hspace{7mm} $\mathrm{rank}\left(
      \begin{bmatrix}
      \boldsymbol{V}\\
      \mathrm{Xops}(i)
      \end{bmatrix} \right) > \mathrm{rank}(\boldsymbol{V})$ and \\%
      \hspace{7mm} $\mathrm{rank}\left(
      \begin{bmatrix}
      \mathrm{validXGauges}\\
      \mathrm{Xops}(i)
      \end{bmatrix} \right) > \mathrm{rank}\left( \mathrm{validXGauges} \right)$}
      
   \State $\mathrm{validXGauges} \gets \mathrm{validXGauges} \cup \mathrm{Xops}(i)$ 
  \EndIf

\EndFor

\If{$\mathrm{validXGauges} == \emptyset$}
\State $w \gets w + 1$
\State Regenerate Xops and Zops
       \If{$w \geq n$}
           \State \textbf{stop}
           \Comment{Algorithm Fails}
       \Else
            \State \textbf{go} to Line 3
        \EndIf
\EndIf

\State $\mathrm{XgChoices} \gets $ NCHOOSEK(validXGauges , gaugesPerStab)

\Comment{List of all combinations of gaugesPerStab elements in\\ \hspace{12mm} validXGauges}

\For{$i$ in $\mathcal{S}_{\rm Xtargets}$}
\For{$j$ in  1:size(XgChoices)}
\State resultantGauge $\gets \boldsymbol{U}(i) + \mathrm{XgChoices}(j)$ (mod 2)
\State residualWeight$(j)$ $\gets$ HammingWeight(resultantGauge)
\EndFor

\State minIndex $\gets$ argmin(residualWeight)

\State $\mathcal{G}_{X} \gets \mathcal{G}_X \cup \mathrm{XgChoices(minIndex)}$


  \If{$\mathrm{size}(\mathcal{G}_X) \geq r$}
     \State Drop all rows from $r+1$ (if they exist)
     \State \textbf{break}
  \EndIf

    
\EndFor

\State Clear the rows $\{ i_1, i_2, \ldots, i_r \}$ and $\{ n+i_1, n+i_2, \ldots, n+i_r \}$ from $\boldsymbol{U}$ to add gauge generators

\State $\boldsymbol{U}([i_1, i_2, \ldots, i_r]) \gets \mathcal{G}_X$

\State Repeat from Line 3 for $\mathrm{Zops}$ by appropriately replacing variables; add $\mathrm{Zops}(i)$ to validZGauges only if HammingWeight$\left( \mathrm{Zops}(i)\, \boldsymbol{\Omega}\, \mathcal{G}_{X}^T \right) = 1$

\Comment{Zops$(i)$ anti-commutes with exactly one $X$-gauge}

\State $\boldsymbol{U}([n+i_1, n+i_2, \ldots, n+i_r]) \gets \mathcal{G}_Z$

\State Replace unused destabilizers in $\mathcal{S'}$ such that $\boldsymbol{U}$ is symplectic
\State \textbf{Return:} $\boldsymbol{U}$, codeDistance($\boldsymbol{U}$)

\end{algorithmic}
\end{algorithm}

\begin{algorithm}[!t]
\caption{\emph{GNarsil} 2: Gauge Operators to Split Stabilizers}
\label{alg:euclid2}

\begin{algorithmic}[1]

\State \textbf{Input:} $\boldsymbol{U} = 
\begin{bmatrix}
    \mathcal{L}_{X} \\
    \mathcal{S} \\
    \mathcal{L}_{Z}\\
    \mathcal{S'}
\end{bmatrix}$, $\{ i_1, i_2, \ldots, i_r \} \subseteq \{ k+1, k+2, \ldots, n \}:$ indices of rows of $\boldsymbol{U}$ that will be removed to add gauge operators

\State \textbf{Initialization:} 



$\mathcal{G}_X \gets \emptyset, \ \mathcal{G}_Z \gets \emptyset$

$\mathrm{validXGauges} \gets \emptyset, \ \mathrm{validZGauges} \gets \emptyset$



$\mathrm{maxSize} \gets $ max. number of gauge candidates to consider; 

$S_{\rm Xtargets} \gets$ indices of rows of $\boldsymbol{U}$ that are $X$-stabilizers; 

$S_{\rm Ztargets} \gets$ indices of rows of $\boldsymbol{U}$ that are $Z$-stabilizers; 

$\mathrm{gaugesPerStab} \gets $ \# of gauge operators per stabilizer; 

$\mathrm{numXGauges}\gets $ $\mathrm{gaugesPerStab}\times\#$ of $X$ stabilizers; 
 
$\mathrm{numZGauges}\gets $ $\mathrm{gaugesPerStab}\times$\# of $Z$ stabilizers; 

Xops $\gets$ weight-$w$ vectors in $\{0,1\}^{n}$ appended with $\boldsymbol{0}$ at 

\hspace{11mm} the end to make length $2n$ (for $X$-gauges); 

Zops $\gets$ weight-$w$ vectors in $\{0,1\}^{n}$ appended with $\boldsymbol{0}$ at 

\hspace{11mm} the beginning to make length $2n$ (for $Z$-gauges); 

$\boldsymbol{V}\gets \boldsymbol{U}(1:(n+k))$ (i.e., remove $\mathcal{S}'$ from $\boldsymbol{U}$) 

\Comment{Find $X$, $Z$ gauge operators}

\For{$i$ in 1:size($X\mathrm{ops}$)}
  
  \If{size(validXGauges) $\geq$ maxSize}
  \State \textbf{break}
  \EndIf
  
  \If{$\mathrm{Xops}(i)\, \boldsymbol{\Omega}\, \boldsymbol{V}^T = \boldsymbol{0}\ (\bmod\ 2)$ and \\ 
  \hspace{7mm} $\mathrm{Xops}(i) \notin  \mathcal{L}_{X}$}

   \State $\mathrm{validXGauges} \gets \mathrm{validXGauges} \cup \mathrm{Xops}(i)$ 
  \EndIf

\EndFor

\If{$\mathrm{validXGauges} == \emptyset$}
\State $w \gets w + 1$
       \If{$w \geq n$}
           \State \textbf{stop}
           \Comment{Algorithm Fails}
       \Else
            \State \textbf{go} to Line 3
        \EndIf
\EndIf

\State $\mathrm{XgChoices} \gets $ NCHOOSEK(validXGauges , gaugesPerStab)

\Comment{List of all combinations of gaugesPerStab elements in\\ \hspace{12mm} validXGauges}

\For{$i$ in $\mathcal{S}_{\rm Xtargets}$}
\For{$j$ in  1:size(XgChoices)}
\State resultantGauge $\gets \boldsymbol{U}(i) + \mathrm{XgChoices}(j)$ (mod 2)
\State residualWeight$(j)$ $\gets$ HammingWeight(resultantGauge)
\EndFor

\State minIndex $\gets$ argmin(residualWeight)

\State $\mathcal{G}_{X} \gets \mathcal{G}_X \cup \mathrm{XgChoices(minIndex)}$


    
\EndFor

\State Remove rows $\{ i_1, i_2, \ldots, i_r \}$ and $\{ n+i_1, n+i_2, \ldots, n+i_r \}$ from $\boldsymbol{U}$; replace with numXGauges and numZGauges empty rows, respectively

\State $\boldsymbol{U}([i_1, i_2, \ldots, i_{\mathrm{numXGauges}}]) \gets \mathcal{G}_X$

\State Repeat from Line 3 for $\mathrm{Zops}$ by appropriately replacing variables; add $\mathrm{Zops}(i)$ to validZGauges only if HammingWeight$\left( \mathrm{Zops}(i)\, \boldsymbol{\Omega}\, \mathcal{G}_{X}^T \right) \geq 1$

\Comment{Zops$(i)$ may  anti-commute with more than one $X$-gauge}

\State $\boldsymbol{U}([n+i_1, n+i_2, \ldots, n+i_{\mathrm{numZGauges}}
]) \gets \mathcal{G}_Z$

\State \textbf{Return:} $\boldsymbol{U}$, codeDistance($\boldsymbol{U}$)

\end{algorithmic}

\end{algorithm}

\section{Examples Found by Our Algorithms}
\label{sec:examples}

\subsection{$\llbr 9,1,4,3 \rrbr$ Bacon-Shor Code}

\begin{figure}
 \includegraphics[scale=0.5]{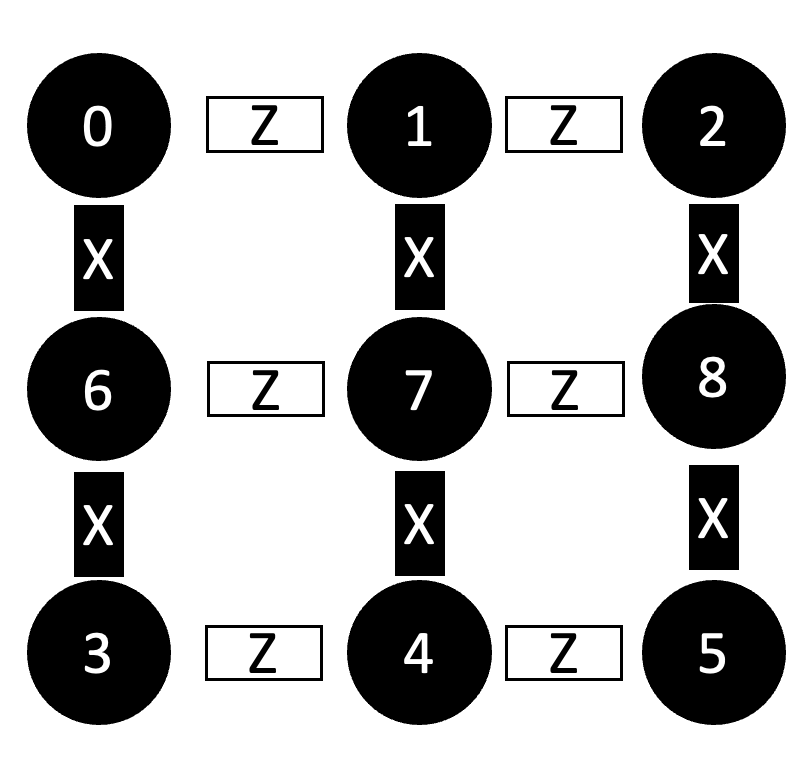}
\centering
\caption{The Bacon-Shor code and its weight-$2$ gauge operators.}
\label{fig:bacon-shor}
\vspace*{-20pt}
\end{figure}

Algorithm \eqref{alg:euclid} was able to find the $\llbr 9,1,4,3 \rrbr $ Bacon-Shor code from the $ \llbr 9,1,3 \rrbr$ Shor (stabilizer) code. 
This required first replacing two of the weight-$2$ $Z$-stabilizers with linearly independent weight-$6$ $Z$-stabilizers from the span of the weight-$2$ $Z$-stabilizers before running the algorithm. 
Finding this prototypical subsystem code example with our algorithm points to its validity. 
We note here that although six gauge operators are shown in Fig.~\ref{fig:bacon-shor}, not all are linearly independent. 
The last two are the product of the linearly independent gauge operators and stabilizers. 
The specific pre-processing of the stabilizers above is a key step that can naturally be generalized to other CSS concatenated codes. 
Specifically, the inner code stabilizers can be multiplied to produce large-weight stabilizers, thereby allowing us to make low-weight gauges from the original (inner code) stabilizers.

\subsection{$\llbr 9,1,2,2 \rrbr$ Rotated Surface Subsystem Code}

\begin{figure}[!b]
\centering
\vspace*{-20pt}
 \includegraphics[scale=0.45]{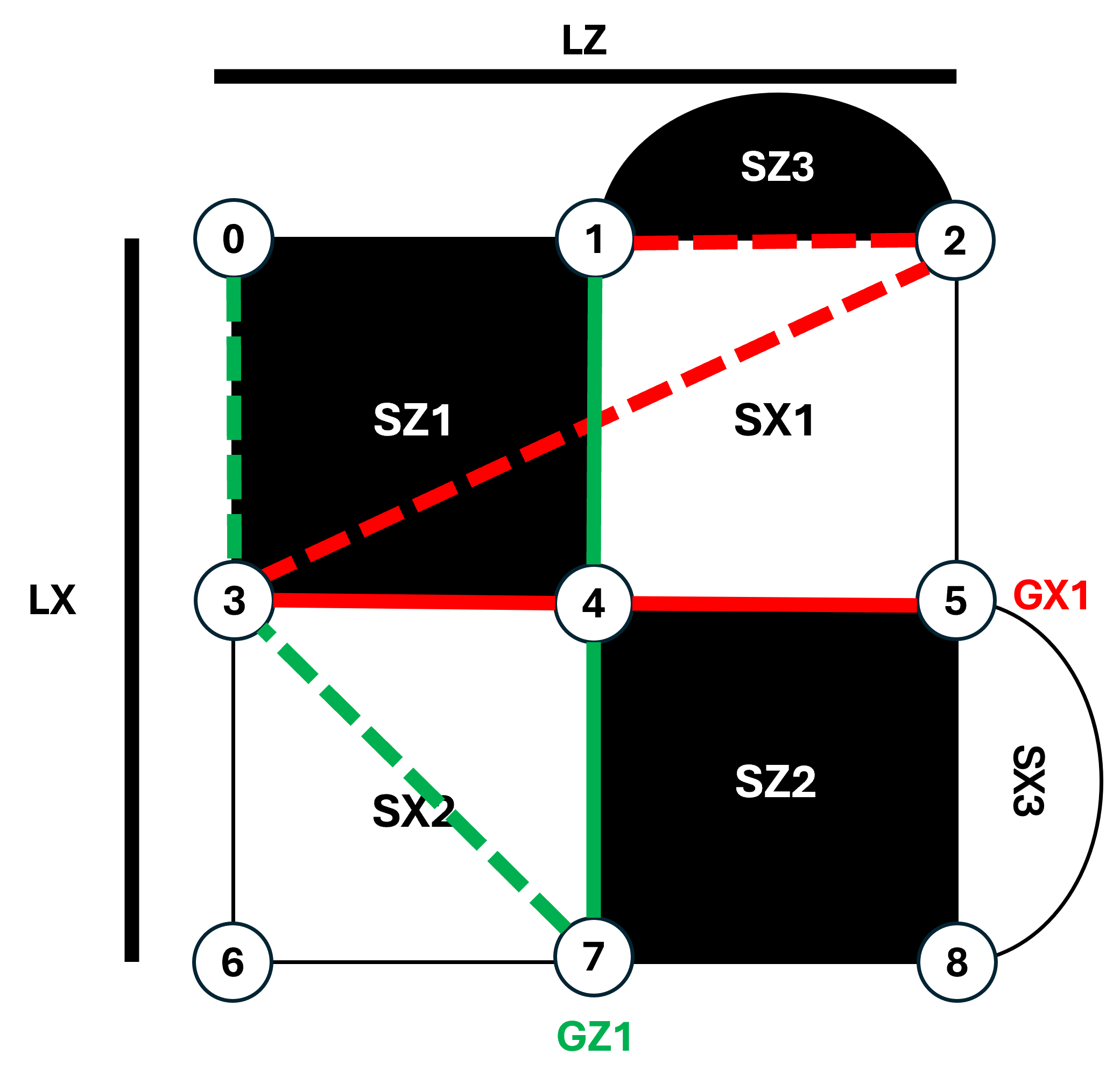}
\caption{The rotated surface subsystem code found by Algorithm~\ref{alg:euclid}. The two red (resp. green) operators are the $X$-gauges (resp. $Z$-gauges), all weight-$3$ (also see Table~\ref{tab:surface_subsystem_gauges}). The product of the pair of red gauge operators, $X_1 X_2 X_3$ and $X_3 X_4 X_5$, gives $SX_{1}$, and the product of the pair of green operators gives $SZ_{1}$. Stabilizers $SX_{2}$, $SZ_{2}$ can be obtained by multiplying $GX_{1}$, $GZ_{1}$ by the respective stabilizers to find the respective dependent gauges. $GX_{2}$, $GZ_{2}$ are not shown here.}
\label{fig:surface_subsystem}
\end{figure}

In this example, we present a novel subsystem version of the $\llbr 9,1,3 \rrbr$ rotated surface code found by Algorithm \eqref{alg:euclid}. 
The code has weight-$4$ stabilizer generators and weight-$3$ gauge operators, as shown in Fig.~\ref{fig:surface_subsystem}. 
We summarize the set of gauge operators, both generators and dependent ones, in Table~\ref{tab:surface_subsystem_gauges}. 
Using the dependent gauges and gauge generators $GX_{1}$, $GZ_{1}$, we can measure all of the weight 4 stabilizers shown above at the cost of the dependent gauge operators being not fully local and a small loss of distance. 
The well-known subsystem version of the surface code in the literature~\cite{higgott2021subsystem,Bravyi-qic12} has weight-$6$ stabilizers instead of the usual weight-$4$ and requires significantly more qubits, e.g., for lattice size $L=3$ the code uses $3L^2 + 4L + 1 = 40$ qubits.
It is not unreasonable to consider the code in Fig.~\ref{fig:surface_subsystem}, but it will be interesting to explore whether there is an intermediate subsystem surface code between these two solutions that still has distance $3$.

\begin{table}[!b]
    \centering
    
    \begin{tabular}{c|c}
    \hline
 \multicolumn{2}{| c |}{List of $ \llbr 9,1,2,2 \rrbr$ subsystem code gauge operators}\\
 \hline
 Gauge Identifier & Gauge Operator \\
 \hline

    $GX_{1}$ & $X_{3}X_{4}X_{5}$ \\ 
     $GX_{2}$ & $X_{3}X_{4}X_{7}$\\ 
     $GX_{1d}$ & $X_{1}X_{2}X_{3}$\\ 
     $GX_{2d}$ & $X_{5}X_{6}X_{7}$\\ 
     $GZ_{1}$ & $Z_{1}Z_{4}Z_{7}$\\ 
     $GZ_{2}$ & $Z_{4}Z_{5}Z_{8}$\\ 
     $GZ_{1d}$& $Z_{0}Z_{3}Z_{7}$\\ 
     $GZ_{2d}$& $Z_{1}Z_{5}Z_{8}$  \\ \hline

    \end{tabular}
    \caption{Gauge operators of the $ \llbr 9,1,2,2 \rrbr$ rotated surface subsystem code. $GX_1, GX_2, GZ_1, GZ_2$ are the independent gauge generators. The letter `d' in subscripts indicates dependent gauge operators that are products of the gauge generators and stabilizers. The stabilizers are shown in Fig.~\ref{fig:surface_subsystem}.}
    \label{tab:surface_subsystem_gauges}
\end{table}

\subsection{$\llbr 100,25,3 \rrbr$ Subsystem Hypergraph Product (SHP) Code}

Using Algorithm \eqref{alg:euclid2}, we present a $\llbr 100,25,3 \rrbr$ SHP code built from a $[10,5]$ linear code~\cite{ryan2004introduction} with parity-check matrix
\begin{equation} \label{eq:H_binary} \boldsymbol{H}=
\begin{bmatrix}
    1&1&1&1&0&0&0&0&0&0 \\
    1&0&0&0&1&1&1&0&0&0 \\
    0&1&0&0&1&0&0&1&1&0 \\
    0&0&1&0&0&1&0&1&0&1\\
    0&0&0&1&0&0&1&0&1&1
\end{bmatrix}.
\end{equation}
We construct the SHP code with the following definitions~\cite{li2020numerical}:
\begin{align}
\label{eq:SHP}
\mathcal{G}_{X} & \coloneqq \left(\boldsymbol{H}\otimes \boldsymbol{I}_{n}\right), \nonumber\\ 
\mathcal{G}_{Z} & \coloneqq \left(\boldsymbol{I}_{n} \otimes \boldsymbol{H} \right), \nonumber\\ 
\mathcal{L}_{X} & \coloneqq \left(\boldsymbol{I}_{n} \otimes \boldsymbol{G}\right),\\ 
\mathcal{L}_{Z} & \coloneqq \left( \boldsymbol{G}\otimes \boldsymbol{I}_{n} \right), \nonumber\\ 
\mathcal{S}_{X} & \coloneqq \left(\boldsymbol{H} \otimes \boldsymbol{G}\right), \nonumber\\ 
\mathcal{S}_{Z} & \coloneqq \left( \boldsymbol{G}\otimes \boldsymbol{H}\right) \nonumber,
\end{align}
where $\boldsymbol{G}$ is the generator matrix of the code defined by the parity check matrix $\boldsymbol{H}$, i.e, $\boldsymbol{G}\boldsymbol{H}^{T}=0$.
Thus, in general, the SHP construction in this form gives an $\llbr n^{2},k^{2},d \rrbr$ code with $n,k,d$ being the respective values for the classical code defined by the parity check matrix $\boldsymbol{H}$.
Here, the key observation is that although the stabilizers for this code are weight-$12$, our algorithm finds a decomposition that only requires weight-$4$ operators with a residual weight of $0$ (resp. $5$) for the $X$-stabilizers (resp. $Z$-stabilizers). 
This is in contrast to the gauge operators found from the closed-form expression in \eqref{eq:SHP}, which are all weight-$4$, requiring residual weights of $6$ and $16$ for $X$- and $Z$-stabilizers respectively. 
Thus, not only does our algorithm find more gauge operators than the closed-form expression, it also finds gauge operators that decompose the given stabilizers more efficiently than the closed-form operators. 
This decomposition makes this high-rate code far easier to implement. 
We also note that high-weight stabilizers are common for the SHP construction, as shown by the $\llbr 49,16,3 \rrbr$ SHP code~\cite{li2020numerical} built from the $[7,4,3]$ Hamming code with stabilizer weights $12$ and $16$, respectively. 
The parity check matrix for the $[7,4,3]$ Hamming code is given below.  
\begin{equation} 
\label{eq:Hamm} 
\boldsymbol{H}=
\begin{bmatrix}
    1&1&1&0&1&0&0\\
    1&1&0&1&0&1&0\\
    1&0&1&1&0&0&1
\end{bmatrix}.
\end{equation}
Hence, our algorithm is likely useful to determine more efficient implementations of SHP codes in general.

\subsection{Subsystem Lifted-Product (SLP) Codes}

As a natural extension of hypergraph product codes, lifted-product (LP) CSS codes~\cite{panteleev2021quantum} are constructed from the hypergraph product of base parity-check matrices $\boldsymbol{A}$ and $\boldsymbol{A}^*$, which is then lifted over the ring $R=\mathbb{F}_{q}(G)$, where $G$ is some group. 
The resulting parity-check matrices are as follows:
\begin{align}
\label{eq:LP}
\boldsymbol{H}_{X}=\left[ \boldsymbol{A} \otimes \boldsymbol{I}, \boldsymbol{I} \otimes \boldsymbol{A} \right ], \nonumber \\
\boldsymbol{H}_{Z}=\left[ \boldsymbol{I} \otimes \boldsymbol{A}^{*},  \boldsymbol{A}^{*}\otimes \boldsymbol{I} \right ]. 
\end{align}
where $\boldsymbol{A}^{*}$ is the conjugate transpose of $\boldsymbol{A}$. 
This construction is typically to obtain good quantum LDPC codes, but its sparsity ends up producing non-local stabilizer operators. 
When LDPC codes are input into~\eqref{eq:LP}, and the resulting $\boldsymbol{H}_{X}, \boldsymbol{H}_{Z}$ are used as input for Algorithm~\ref{alg:euclid2}, the algorithm is able to find several gauge operators for these codes. 
However, none of them are useful since the residual weight will greatly exceed the weight of the stabilizers themselves. 
This fact seems to stem from the sparse nature of the stabilizers: in order to commute with them, the gauge operators must also be sparse, which in turn requires a larger amount of residual weight to cover the spread of the stabilizers. 
The precise understanding of this situation for LP codes is an interesting future direction.

\textbf{The SLP Construction:}
To find a lifted code with a better gauge decomposition, one is tempted to use the SHP construction in~\eqref{eq:SHP} as a natural extension of LP codes. 
We will refer to this as the SLP construction. 
By employing the base parity check-matrix $\boldsymbol{A}$ along with its (base) generator matrix $\boldsymbol{G}_{\boldsymbol{A}}$ into~\eqref{eq:SHP} and then lifting the construction by a circulant of size $L$, we obtain an $\llbr Ln^{2},Lk^{2},D \rrbr$ subsystem code, where $n,k$ are the respective values defined by the base parity-check matrix $\boldsymbol{A}$. 
Here, $\boldsymbol{A}\boldsymbol{G}_{\boldsymbol{A}}^*=0$, which implies that $\mathcal{G}_{X}\mathcal{S}^*_{Z}=0$. 
This lifting procedure yields codes with potentially larger distances than a typical SHP code. 

\subsubsection{Comparing the SHP and SLP Constructions}
For the most direct comparison, we choose a binary base matrix:
\begin{equation} 
\label{eq:Abinary} 
\boldsymbol{A}=
\begin{bmatrix}
    0&1&1\\
    1&1&0
\end{bmatrix},
\end{equation}
where the generator matrix of \eqref{eq:Abinary} is given by:
\begin{equation} 
\label{eq:Abinarygen}
\boldsymbol{G}_{\boldsymbol{A}}=
\begin{bmatrix}
    1&1&1
\end{bmatrix}.
\end{equation}
Placing into \eqref{eq:SHP}, this yields a $\llbr 9,1,3 \rrbr$ SHP code, which is simply the Bacon-Shor code. 
For the SLP construction, we interpret~\eqref{eq:Abinary} as the matrix of powers of monomials corresponding to $L=2$ circulant matrices. 
Thus, we define our new base matrix as:
\begin{equation} 
\label{eq:Acirc} 
\boldsymbol{B}=
\begin{bmatrix}
    1&x&x\\
    x&x&1
\end{bmatrix}.
\end{equation}
Clearly, Eqn.~\eqref{eq:Abinarygen} interpreted in the same fashion is no longer a valid solution, so we must use another generator matrix. It can be shown that the following matrix is a valid generator matrix for the base matrix $\boldsymbol{B}$:
\begin{equation} 
\label{eq:Acirc} 
\boldsymbol{G}_{\boldsymbol{B}}=
\begin{bmatrix}
    1+x&1+x&0\\
    1+x&0&1+x\\
    x&0&1
\end{bmatrix}.
\end{equation}
This yields a  $\llbr 18,2,2 \rrbr$ SLP code. The SLP code here has the same rate as the SHP code but with more physical qubits and a small loss in distance. However, we will see other cases where the distance fairs favorably compared to an LP code with the same base matrix.
We now look at a few non-trivial examples of the SLP construction and their performance.

\subsubsection{$\llbr 27,12,2 \rrbr$ SLP code}

Given $L=3$, the $\llbr 27,12,2 \rrbr$ SLP code can be constructed by the following base matrix:
\begin{equation} 
\label{eq:A_27} 
\boldsymbol{A}=
\begin{bmatrix}
    1+x+x^2&1+x&x
\end{bmatrix},
\end{equation}
along with the base generator matrix of the code:
\begin{equation} \label{eq:GA_27} \boldsymbol{G}_{\boldsymbol{A}}=
\begin{bmatrix}
   x^2&x&1\\
   x&x^2&x\\
   1&0&1+x+x^2
\end{bmatrix}.  
\end{equation}

After inserting these into \eqref{eq:SHP} and lifting the construction, we find that the Pauli weight of the code's stabilizers is $18$, and the Pauli weight of the gauge operators is $6$. 
We also find that the gauge operators in the closed-form construction decompose the stabilizers with a residual weight of $9$, given three weight-$6$ gauge operators as input. 
However, our algorithm is able to decompose the stabilizers with residual weights of $4,6,8$ for the $X$- and $Z$-stabilizers, an improvement over the closed-form gauge operator decomposition. 
Finally, we also note that the rate of the SLP construction is $\frac{4}{9}$, which is greater than the rate of the $\llbr39,12,2\rrbr$  LP code constructed from the same base matrix, $\frac{4}{13}$. 
Note that this is also while preserving the distance of the code.

\subsubsection{$\llbr 775,124,20 \rrbr$ SLP code}
For our final example, we give the SLP code constructed by Tanner's $(3,5)$ QC-LDPC code with $L=31$. 
Here, we use the following base matrix~\cite{raveendran2022finite}:
\begin{equation} 
\label{eq:B31} 
\boldsymbol{B}=
\begin{bmatrix}
    x&x^{2}&x^{4}&x^{8}&x^{16}\\
    x^{5}&x^{10}&x^{20}&x^{9}&x^{18}\\
    x^{25}&x^{19}&x^{7}&x^{14}&x^{28}
\end{bmatrix}. 
\end{equation}
As shown by Smarandache, and then Chimal-Dzul, Lieb, and Rosenthal \cite{smarandache2022using},              \cite{chimal2022generator}, a generator matrix for \eqref{eq:B31} can be constructed using the matrix:
\begin{align} 
\boldsymbol{G}_{\boldsymbol{B}} &=
\begin{bmatrix}
    u_{11}&u_{12}&u_{13}&u_{14}&0\\
    u_{21}&u_{22}&u_{23}&0&u_{25}\\
    f&f&0&0&0\\
    f&0&f&0&0\\
    f&0&0&f&0\\
     f&0&0&0&f
\end{bmatrix}, \nonumber \\
\label{eq:partsofGfull31}
u_{11} &= x^{28} + x^{25} + x^{18} + x^{16} + x^{5} + x, \nonumber \\
u_{12} &= x^{23} + x^{22} + x^{20} + x^{17} + x^{7} + x^{4},\nonumber\\
u_{13} &= x^{29} + x^{25} + x^{21} + x^{12} + x^{5} + x,\nonumber\\
u_{14} &= u_{25} = x^{28} + x^{18} + x^{16} + x^{14} + x^{9} + x^{8},\nonumber\\
u_{21} &= x^{27} + x^{24} + x^{19} + x^{11} + x^{10} + x^{2},\nonumber\\
u_{22} &= x^{30} + x^{28} + x^{26} + x^{18} + x^{16} + x^{6},\nonumber\\
u_{23} &= x^{20} + x^{14} + x^{9} + x^{8} + x^{7} + x^{4},\nonumber\\
f &= \frac{x^{31}-1}{x-1} = x^{30} + \cdots + x + 1.
\end{align}

We find that the code has stabilizer weights of $120$, $310$, and $465$ and gauge operator weights of $5$. 
Due to the large number of physical qubits, this code is beyond the practical scale that our algorithm can handle. 
Our algorithm is suited for small- to medium-sized codes due to its exponential memory complexity, but this decomposition by GNarsil may be achievable with a sufficiently large amount of memory. 
Our SLP example demonstrates that the SLP construction can be quite advantageous if a generator matrix can be found for a certain base matrix. 
For the same base matrix~\eqref{eq:B31}, an LP($\boldsymbol{B}$,$\boldsymbol{B}^*$) construction yields a $\llbr 1054,124,20 \rrbr$ code, which trails the SLP version both in rate. 
If a circulant form generator matrix is found for a base matrix, then the resulting SLP code will have a higher rate than its LP counterpart with the same code dimension. 
No relationship between the distance of the two constructions has been formalized, but it seems that the SLP code will at least have the same distance as its LP counterpart.
This makes the SLP construction for a given base matrix a very attractive construction whenever possible.

\section{Conclusion}
\label{sec:conclusion}

In this paper, we have introduced a new set of algorithms, which we call \emph{GNarsil}, for deriving subsystem codes from an input ``seed'' stabilizer code. 
We have demonstrated that these algorithms not only recover well-known examples of subsystem codes but also find interesting new ones, such as a novel version of the rotated surface subsystem code and a new SHP code that is more efficient than the closed-form construction in~\cite{li2020numerical}. 
We also reported that we could not find useful gauge decompositions of LP codes due to the highly non-local structure of the stabilizers. 
However, by using our new SLP construction, we can construct codes with excellent parameters that also have promising stabilizer decomposition properties, which our algorithms can improve. 
As noted, this also depends on finding a generator matrix for the base matrix in circulant form, which is not always guaranteed to exist. 

We foresee these algorithms becoming useful tools for finding subsystem versions of stabilizer codes that may prove easier to implement due to the stabilizers' decomposition into smaller-weight gauge operators. 
However, it is clear that this problem remains complex, as the solution spaces for useful gauge generators are sparse, making it likely that our algorithms are optimal for this case, even though they are exponentially complex in memory. 
Thus, our algorithms work best for small- to medium-sized codes. 
Improving our algorithms' performance will most likely require specializing them for certain code constructions. 
This would allow us to exploit code symmetries to find optimal gauge operators. 

We also foresee that the measure of residual weight can be useful for understanding the properties of good subsystem codes and the symmetries that they may possess. 
Finally, we wish to extend this work into looking at the relationship between fault-tolerant operations on a stabilizer code and its derived subsystem code(s) by extending tools such as the LCS algorithm~\cite{Rengaswamy-tqe20}. 
This can potentially be approached by observing how new gauge operators found by our algorithm change the structure of logical gates for our code from the seed stabilizer code. 
Insights into this relationship may be useful for the development of logical operations on Floquet codes~\cite{Hastings-quantum21}, at least the ones with parent subsystem codes~\cite{Davydova-prxq23}.

 \section*{Acknowledgements}
 We thank Nithin Raveendran and Bane Vasić at the University of Arizona for their insights into LP and quasi-cyclic (QC) LDPC codes, respectively, especially about finding circulant form generator matrices for QC-LDPC codes. 
 Finally, we would like to thank M. Sohaib Alam of the Quantum Artificial Intelligence Laboratory at NASA Ames Research Center for his insights on subsystem codes and our algorithms. 

 The work of N.~R. was partially supported by the National Science Foundation under Grant no. 2106189.

\bibliographystyle{IEEEtran}
\bibliography{cites}

\appendices

\section{Proof of Algorithm's ability to find all possible representations of a subsystem code using Binary Symplectic Matrices}
 \label{sec:AppendixA}

\begin{theorem} Given an $\llbr n,k,d\rrbr$ CSS code with logical operators $\boldsymbol{X_{i}},\boldsymbol{Z_{i}}$ for each logical qubit, the GNarsil algorithms can find all representations for the $2r$ gauge generators of the derived $\llbr n,k,r,d \rrbr$ subsystem code, where $2n-2r$ of the rows of the matrix $\boldsymbol{U}$ representing the code~\eqref{eq:U_matrix} are fixed.

\begin{IEEEproof}
\normalfont
Let $r$ be the number of gauge qubits chosen from the input code. 
The input code is described by a $2n \times 2n$ matrix $\boldsymbol{U}$, a symplectic matrix that describes the code's logical operators and stabilizers. 
The row vectors $\{ \boldsymbol{u}_{1}, \ldots, \boldsymbol{u}_{2n} \}$ of $\boldsymbol{U}$ then span the space $\mathcal{U}	= \mathbb{F}^{2n}_{2}$. 
We choose $2r$ vectors $\{ \boldsymbol{u}_i : i\in \mathcal{I} \}$,  
$\mathcal{I} = \{ i_1, i_2, \ldots, i_r, n+i_1, n+i_2, \ldots, n+i_r \}$, from $\boldsymbol{U}$ which correspond to a submatrix $\boldsymbol{U}_r$ of $\boldsymbol{U}$. 
These vectors from $\mathcal{I}$ form a symplectic basis for a $2^{2r}$-dimensional subspace $\mathcal{U}_{r} \subseteq \mathcal{U}$.
Since the subspace is symplectic, we see that any vector $\boldsymbol{u}_{i}, i\in \mathcal{I}$, has a symplectic product $\langle \boldsymbol{u}_{i},\boldsymbol{u}_{j}\rangle_{s}=0 $ (mod 2) with any vector $\boldsymbol{u}_{j}, j\notin \mathcal{I}$.  

We now replace the vectors from  $\boldsymbol{U}_r$ with vectors $\tilde{\boldsymbol{u}} \in \mathcal{U}_r$ such that $\langle \tilde{\boldsymbol{u}}_{i}, \tilde{\boldsymbol{u}}_{i+r}\rangle_{s} = 1\ (\bmod\ 2)$. 
To see which pairs of vectors in $\mathcal{U}_r$ form valid symplectic pairs we arrange all of the symplectic products of vectors in $\mathcal{U}_r$ into a matrix $\boldsymbol{P}=\sum_{ij} \langle \boldsymbol{u'}_{i},\boldsymbol{u'}_{j}\rangle_{s}\ket{i}\bra{j} $ (mod $2$), where $\boldsymbol{u'}_{i},\boldsymbol{u'}_{j} \in \mathcal{U}_r$. 
We use $\boldsymbol{P}$ to count the number of possible representations generated of symplectic pairs of $\mathcal{U}_{r}$. 
The number of representations is defined as the number of unique matrices $\boldsymbol{V}$ built from $\boldsymbol{U}$ by replacing the vectors $\{ \boldsymbol{u}_i : i\in \mathcal{I}\}$ with vectors $\tilde{\boldsymbol{u}}$  such that $\boldsymbol{V}\boldsymbol{\Omega}\boldsymbol{V}^T=\boldsymbol{\Omega}$, where $\boldsymbol{\Omega}=
\begin{bsmallmatrix} 
  \boldsymbol{0} & \boldsymbol{I}_n \\
  \boldsymbol{I}_n & \boldsymbol{0}
\end{bsmallmatrix}$.
  
This condition is equivalent to all constraints needed to be satisfied by a valid subsystem code. The number of representations for a given $r$ can be found by examining $\boldsymbol{P}$. 
For instance, in the case of $r=2$ we see that:
\begin{equation}
\boldsymbol{P} = 
\begin{bmatrix}
\begin{smallmatrix}
       0 & 0 & 0 & 0 & 0 & 0 & 0 & 0 & 0 & 0 & 0 & 0 & 0 & 0 & 0 & 0\\
       0 & 0 & 0 & 0 & 1 & 1 & 1 & 1 & 0 & 0 & 0 & 0 & 1 & 1 & 1 & 1\\
       0 & 0 & 0 & 0 & 0 & 0 & 0 & 0 & 1 & 1 & 1 & 1 & 1 & 1 & 1 & 1\\
       0 & 0 & 0 & 0 & 1 & 1 & 1 & 1 & 1 & 1 & 1 & 1 & 0 & 0 & 0 & 0\\
       0 & 1 & 0 & 1 & 0 & 1 & 0 & 1 & 0 & 1 & 0 & 1 & 0 & 1 & 0 & 1\\
       0 & 1 & 0 & 1 & 1 & 0 & 1 & 0 & 0 & 1 & 0 & 1 & 1 & 0 & 1 & 0\\
       0 & 1 & 0 & 1 & 0 & 1 & 0 & 1 & 1 & 0 & 1 & 0 & 1 & 0 & 1 & 0\\
       0 & 1 & 0 & 1 & 1 & 0 & 1 & 0 & 1 & 0 & 1 & 0 & 0 & 1 & 0 & 1\\
       0 & 0 & 1 & 1 & 0 & 0 & 1 & 1 & 0 & 0 & 1 & 1 & 0 & 0 & 1 & 1\\
       0 & 0 & 1 & 1 & 1 & 1 & 0 & 0 & 0 & 0 & 1 & 1 & 1 & 1 & 0 & 0\\
       0 & 0 & 1 & 1 & 0 & 0 & 1 & 1 & 1 & 1 & 0 & 0 & 1 & 1 & 0 & 0\\
       0 & 0 & 1 & 1 & 1 & 1 & 0 & 0 & 1 & 1 & 0 & 0 & 0 & 0 & 1 & 1\\
       0 & 1 & 1 & 0 & 0 & 1 & 1 & 0 & 0 & 1 & 1 & 0 & 0 & 1 & 1 & 0\\
       0 & 1 & 1 & 0 & 1 & 0 & 0 & 1 & 0 & 1 & 1 & 0 & 1 & 0 & 0 & 1\\
       0 & 1 & 1 & 0 & 0 & 1 & 1 & 0 & 1 & 0 & 0 & 1 & 1 & 0 & 0 & 1\\
       0 & 1 & 1 & 0 & 1 & 0 & 0 & 1 & 1 & 0 & 0 & 1 & 0 & 1 & 1 & 0   
\end{smallmatrix}
\end{bmatrix}.
\end{equation}
By examining $\boldsymbol{P}$, we first see that there are $2^{2r}-1 = 15$ non-zero vectors of $\mathcal{U}_r$ that can be chosen as the first vector. 
Each of these vectors has $2^{2r-1}=8$ available symplectic partners. 
For finding the third vector to add to $\boldsymbol{V}$, out of $2r$ needed vectors, we look for possible candidates by searching for the columns of $\boldsymbol{P}$ where the first two vectors' rows are $0$, i.e., the third vector is symplectically orthogonal to the first two vectors added to $\boldsymbol{V}$.
We find $4$ such columns but note that one is the column corresponding to the trivial zero vector. 
Therefore, we have $2^{2r-2}-1=3$ possible nontrivial choices for the third vector. 
Finally, we search for the compatible symplectic pairs of the third vector, which gives us $2^{2r-3}=2$. 
Thus, our total number of subsystem codes for $r=2$ is $720$. 

However, this number includes the multiplicity of codes due to the possible permutations of the $2r$ rows that still allow $\boldsymbol{V}$ to satisfy  $\boldsymbol{V}\boldsymbol{\Omega}\boldsymbol{V}^T=\boldsymbol{\Omega}$. 
By looking at all possible permutations of the rows such that we preserve the symplectic pairs between the row vectors, we find that the multiplicity is $8$ for $r=2$. 
Thus, the total number of unique representations for $r=2$ is $90$. 
We can extrapolate this procedure to arbitrary $r$ and find that the number of representations for a given $r \geq 2$ is
\begin{equation} 
\frac{1}{\mathcal{M}} \left( \prod_{l\in \{ 0,1,2, \ldots, r-1 \}} 2^{2r-2l}-1 \right) \left( \prod_{m \in \{1,3,5, \ldots ,2r-1 \}} 2^{2r-m} \right),
\end{equation} 
where $\mathcal{M}$ is the multiplicity of the $2r$ rows.
We conclude that these form all possible representations of the subsystem code as the gauge generators must necessarily come from $\mathcal{U}_r$.
\end{IEEEproof}
\end{theorem}

\end{document}